\documentclass[prl,twocolumn,aps]{revtex4}
\usepackage{graphicx}
\usepackage{natbib}
\usepackage{color}
\usepackage{units}
\usepackage{amsmath, amsthm, amssymb}
\usepackage[normalem]{ulem}

\def\strutdepth{\dp\strutbox}
\def\nw#1{\strut\vadjust{\kern-\strutdepth\vtop to0pt{\vss\hbox to\hsize
{\hskip\hsize\hskip5pt$\leftarrow$\hss\strut}}}{\em #1}}
\begin{document}

\title{Two-component self-contracted droplets:\\ long-range attraction and confinement effects}

\author{Adrien Benusiglio$^{1}$, Nate Cira$^{1}$, Anna Wei Lai$^{1}$ and Manu Prakash$^{1}$}

\affiliation{
$^1$Department of Bioengineering, 
Stanford University, 
450 Serra Mall, California 94305, USA.\\}

\begin{abstract}
{Marangoni self-contracted droplets are formed by a mixture of two liquids, one of larger surface tension and larger evaporation rate than the other. Due to evaporation, the droplets contract to a stable contact angle instead of spreading on a wetting substrate. This gives them unique properties, including absence of pinning force and ability to move under vapor gradients, self- and externally imposed. We first model the dynamics of attraction in an unconfined geometry and then study the effects of confinement on the attraction range and dynamics, going from minimal confinement (vertical boundary), to medium confinement (2-D vapor diffusion) and eventually strong confinement (1-D).  "Self-induced" motion is observed when single droplets are placed close to a vapor boundary toward which they are attracted, the boundary acting as an image droplet with respect to itself. When two droplets are confined between two horizontal plates, they interact at a longer distance with modified dynamics. Finally, confining the droplet in a tunnel, the range of attraction is greatly enhanced, as the droplet moves all the way up the tunnel when an external humidity gradient is imposed. "Self-induced" motion is also observed, as the droplet can move by itself towards the center of the tunnel. Confinement greatly increase the range at which droplets interact as well as their lifetime and thus greatly expands the control and design possibilities for applications offered by self-contracted droplets.}
\end{abstract}

\date{\today}%

\maketitle

Here we study the motion of two component droplets that do not suffer from pinning and move due to vapor gradients, as studied experimentally \cite{Cira-2015}. Our group previously showed that droplets of the right two components placed on a high energy surface contract due to evaporation at the origin of a Marangoni flow, forming self-contracting Marangoni droplets (one component must have a larger surface tension and larger evaporation rate than the other, and the mixture should wet the substrate without evaporation). When two droplets are placed close by, they attract each other. We showed that this motion was due to vapor gradients. Previous work showed that it was possible to make regular sessile droplets move by embedding energy gradients in the solid surfaces \cite{Brochard-1989, ondarccuhu1992etalement, Chaudhury-1992}, or by modification of the surface by the droplet \cite{bain1994rapid, dos1995free, shanahan1997start}, methods that have the inconvenience of either having a non-editable or non-repeatable trajectory. Other work showed that droplets could move under gradients of temperature \cite{brzoska1993motions, valentino2003thermocapillary, gomba2010regimes} or modification of the surface tension with electric fields \cite{abdelgawad2008all, choi2012digital}, methods that couldn't get rid of the inherent pinning of the droplet and need specific integration of electronics in the substrate, although the use of lubricant-impregnated surfaces greatly reduces pinning and greatly increases the speed of droplets in thermocapillary motion \cite{bjelobrk2016thermocapillary}. The Marangoni self-contracted droplets do no need a surface more specific than being wetted by both components, and moreover, because they move under the influence of each other, are able to autonomously accomplish operations. In this letter we study the dynamics of attraction of two droplets together. We present various experiments and models in which we increase the confinement from minimal to strong, and show how the dynamics and range of attraction can be controlled and greatly enhanced. 

\begin{figure*}
\includegraphics[width=2\columnwidth]{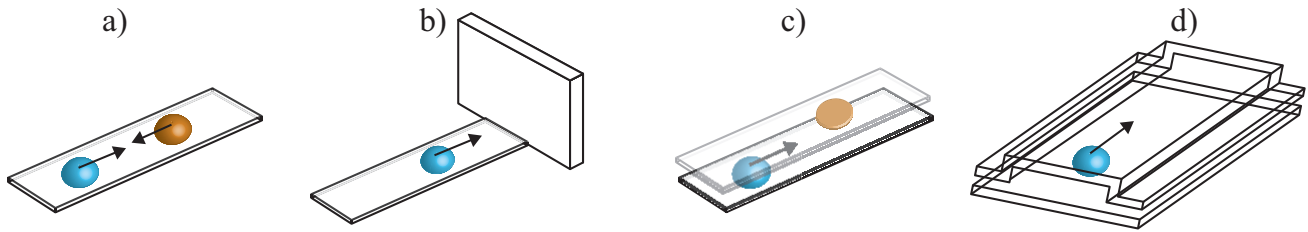}
\caption{Confinement of the droplet, (a) open half space, (b) wall, (c) 2D, (d) 1D (the droplet size is exaggerated for clarity).\label{fig: 3D:2D:1D} }

\end{figure*}

\section{Long-range attraction}

\subsection{Attraction and chasing}
We use a mixture of propylene glycol and water, to form droplets deposited on clean glass slides (\textit{Materials and Methods}). The more volatile component is water, and droplets move in response to humidity gradients. When two millimeter-size droplets are deposited less than three diameters apart, they move toward each other until they contact. If the self-contracted droplets have the same concentration, they then merge together. If the droplets have a difference of concentration large enough to delay coalescence, they then enter a chasing phase, where the droplet of lower surface tension chases the droplet of larger surface tension for up to several minutes \cite{karpitschka2014sharp, karpitschka2010quantitative, karpitschka2012noncoalescence, Cira-2015}. A similar chase can happen with droplets that are not Marangoni-contracted on clean surfaces \cite{sellier2011self}, except that the chasing droplet gets elongated into a strip. In this letter we will study the phase of long range attraction between droplets, and the influence of confinement in various geometries.

\subsection{Observations}
\begin{figure}
\vspace{3cm}
\includegraphics[width=\columnwidth]{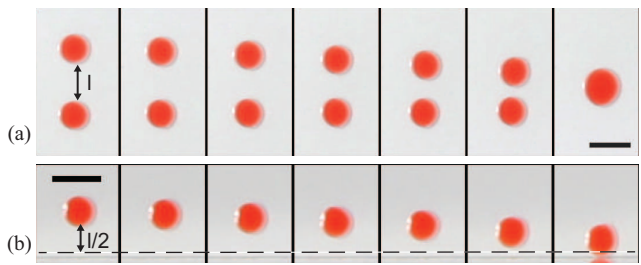}\\
\caption{(a) Two 10 $\%$ PG, $V= 0.5$ $\mu$L droplets attract each other. (b) One 10 $\%$PG, $V=0.5$ $\mu$L moves towards a vertical wall ( position is highlighted by a dashed line). The reflection of the droplet into the wall is visible in the last image. The images are spaces by 1.5 s, the solid line represents 4 mm. \label{fig:chrono3D}}
\end{figure}

\begin{figure}
\includegraphics[width=\columnwidth]{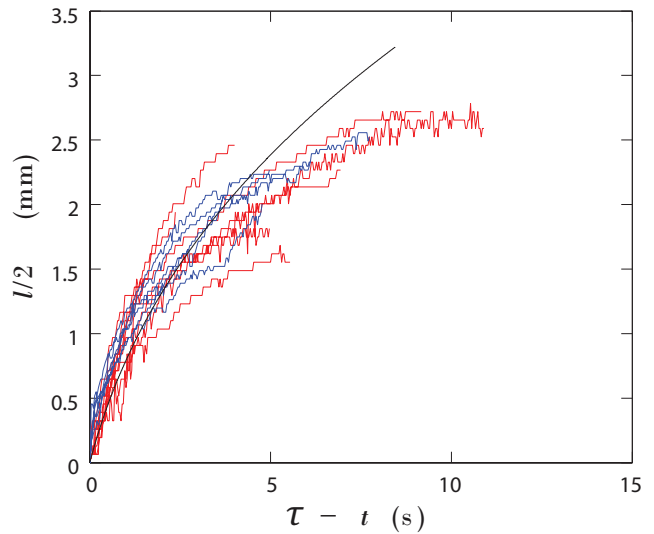}
\caption{Dynamic of attraction between two identical droplets without confinement and of one droplet towards a wall. Blue: multiple trajectories of two 10 $\%$ PG, $V=0.5$ $\mu$L droplets coming together, we plot half of the distance between the droplets as a function of ($\tau-t$) where $\tau$ is the time of contact. Red: multiple trajectories of one droplet of same characteristics moving in the direction of a wall, we plot the distance between the droplet and the wall as a function of ($\tau-t$). Solid line: prediction of the model. \label{fig: 3Dattraction}}
\end{figure}

Fig. \ref{fig:chrono3D} (a) shows a chronophotography of two identical droplets placed on a glass slide without any other confinement [Fig. \ref{fig: 3D:2D:1D} (a)]. They move towards each other at an increasing velocity and eventually merge together. We show the distance between the droplets as a function of time for several repeated experiments in Fig. \ref{fig: 3Dattraction}. $l$ is the shortest border to border distance between the droplets and the time is rescaled as $\tau -t$, with $\tau$ the time of contact ($l=0$). We observe a good repeatability of the experiment for $l$ smaller than 2 droplets diameters.

To discuss the dynamic of attraction between the droplets, we first have to discuss the influence of humidity on the droplets.

\subsection{Droplet contact angle as a function of humidity}
When a single droplet is placed at a controlled humidity, we observe that its contact angle decreases as the humidity is increased. For the droplets of 10$\%$ and 30$\%$ PG volume concentration, the cosine of the contact angle is a linear function of the humidity over a large range of humidity:
\begin{equation}
\cos(\theta)(RH)=m \times RH +b,
\label{eq:theta_RH}
\end{equation}

with $m$ the slope and $b$ a constant, $m$= 2.5 10$^{-4}$ for 0.5 $\mu$L 10$\%$ PG droplets and 3.23 10$^{-4}$ for 0.5 $\mu$L 30$\%$ PG droplets, $RH$ taking values between 0 and 100. The droplet is constantly evaporating water (water is 100 times more volatile than PG), and because it is evaporating faster at the border \cite{hu2002evaporation, eggers2010nonlocal} which is thinner, the border gets enriched in PG. As the PG/water mixture surface tension is monotonically decreasing with PG concentration \cite{karpitschka2010quantitative}, the border has a smaller surface tension. As a result a Marangoni flow from the border to the apex stops the spreading of the drop and it adopts a constant apparent contact angle $\theta$. When the humidity is increased, evaporation is reduced and $\theta$ decreases.
The value of $\theta$ can be interpreted as a the resultant of the horizontal force balance between the bulk of the droplet, of surface tension $\gamma_\mathrm{bulk}$ and its border of surface tension $\gamma_\mathrm{film}$, through the relation\cite{Cira-2015}:
\begin{equation}
\gamma_\mathrm{bulk} \cos(\theta)=\gamma_\mathrm{film}.
\label{equ:modYoung}
\end{equation}

\subsection{Attraction force}
When the droplet is placed in an external humidity gradient, the concentration of its border is now a local function of the rate of evaporation, itself a function of the local humidity gradient, when the bulk concentration remains quasi-constant. If the concentration is uneven around the droplet, one side pulls more than the other, and the droplet moves in response. We don't have direct measurement of the concentration as a function of the local gradient, but only of $\theta$ as a function of a uniform external humidity.

To estimate the force a droplet feels in non-uniform humidity, we make the simplifying assumption that the humidity field imposed by other sources than the droplet in not perturbed by the droplet itself, and that at the location $x$ where an external source imposes a value $RH=RH_\mathrm{ext}(x)$, the droplet's border concentration is the one it would have in a uniform external humidity of the same value. We then estimate the local surface tension of the border from Eq. \ref{equ:modYoung}. Integrating the surface tension along the perimeter of the droplet, we obtain the net attraction force $F_a$ acting on the droplet:
\begin{equation}
F_a=2\int_0^\pi \gamma_\mathrm{bulk} \cos \left[ \theta(RH_\mathrm{ext}(x)) \right] R d\theta .
\end{equation}
Introducing Eq. \ref{eq:theta_RH} we obtain:
\begin{equation}
F_a=2\int_0^\pi \gamma_\mathrm{bulk} m \times RH_\mathrm{ext}(x) R d\theta .
\label{equ:Fattract}
\end{equation}
If the humidity imposed by the surroundings of the mobile droplet is uniform, the net force is equal to zero and the droplets stays motionless. If the mobile droplet is placed close to another droplet (attracting droplet), the humidity is larger between the two drops, and $F_a$ is directed towards the attracting droplet.\\

The 10$\%$ PG droplets, viscosity $\eta=4. 10^{-3}$ Pa/s, density $\rho=1000$ kg/m$^3$, radius $R=1$ mm typically move at the velocity $U=1$ mm/s so that the Reynolds number $Re=\frac{\rho R U}{\eta}$ is of order 1. In the low Reynolds number limit we neglect inertia, and the velocity of the droplet can be estimated from the balance between $F_a$ and a resistive force due to viscous dissipation during motion. The viscous dissipation takes place at the perimeter of the moving droplet where the gradient of velocity is the strongest. From measurements of the velocity of droplets moving down a slope, we observed that the resistive force is a linear function of the droplet velocity: $F_{d}=U/C$, with a coefficient of proportionality $C$ expressed in m.s$^{-1}$.N$^{-1}$. When an attractive force is applied to a droplet, it should then move at the velocity:

\begin{equation}
U=C \times F_a.
\label{eq:U_F}
\end{equation}
To model the attraction between two droplets, the last step is to estimate the gradient of humidity one droplet imposes on the other.

\section{No confinement / Vertical wall Confinement}
\label{sec:3Dmodel}
\subsection{Two droplets attraction}
In the situation where the droplet evaporation is not confined except by the substrate on which it sits [Fig. \ref{fig: 3D:2D:1D} (a)], we will model the humidity field around one droplet as the one formed by a spherical source in an infinite space. The diffusion equation gives, with the two conditions at the limit being that $RH$ is saturated at the surface of the droplet and $RH=RH_0$ at infinity:
\begin{equation}
RH(r)=\frac{(1-RH_0)R}{r}+RH_0,
\label{equ:3Dvap}
\end{equation}
with $r$ the radial distance to the center of the droplet.
The gradient of humidity is large at less than one radius from the droplet and quickly vanishes at a couple radii, what is qualitatively in agreement with the observed motion of two droplets one towards the other [Fig. \ref{fig:chrono3D} (a)].
Replacing Eq. \ref{equ:3Dvap} into Eq. \ref{equ:Fattract} we numerically calculate $F_a$, deduce $U$ from Eq. \ref{eq:U_F}, and finally obtain $l(t)$ by direct integration. We compare our estimation with experiments in Fig. \ref{fig: 3Dattraction}. We observe that the dynamics of attraction are well captured by the model.

\subsection{Wall}
When two identical droplets attract each other, the vertical plane between the droplets is a plane of symmetry and there is no diffusion of humidity through the plane. The plane of symmetry is then analogous to a diffusion barrier. To test this hypothesis, we placed a vertical piece of acrylic 7x7 cm width by height against a treated glass slide, as represented in Fig. \ref{fig: 3D:2D:1D} (b). We observe in Fig. \ref{fig:chrono3D} (b) that when a droplet is placed at two to three diameters from the wall, it moves towards the wall on the shortest path. We compare the dynamics of two identical droplets attracting each other and of one droplet moving towards a vertical wall in Fig. \ref{fig: 3Dattraction}, and observe that they are identical. As expected a wall acts like it is equivalent to a droplet, image of the first one as respect to the wall. 
This idea is very fundamental for the rest of the article: confinement of vapor diffusion modifies the dynamics of attraction between droplets, and the interaction of a single droplet with a confining geometry is sufficient to cause motion.

\section{Two-dimensional confinement}

\begin{figure}
\includegraphics[width=\columnwidth]{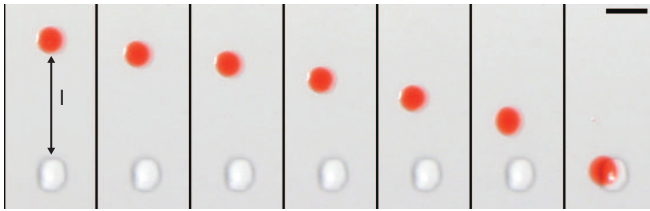}
\caption{2-dimensional confinement. The 10$\%$ PG droplet at the top of the image ( placed on the top surface of the bottom slide) moves towards a pure water droplet at the bottom of the image (placed on the bottom side of the top slide) until they align, the images are spaced by 3 s and the solid line represents 4 mm. \label{fig:2Dmontage}}
\end{figure}

\subsection{Purpose and experimental setup}
Cira \textit{et al} showed that to attract each other, two droplets do not have to be on a continuous surface. Droplets placed on two facing horizontal plates at a millimeter distance will also attract each-other until they align one above the other. 

We now investigate this situation of confinement. The droplets are placed on two facing glass slides (75x25 mm) separated by a 1 mm gap. The bottom slide is cleaned and receives a 0.5 $\mu$L mobile droplet of 10$\%$ PG. The top slide is not cleaned and receives a 1 $\mu$L droplet of pure water that is pinned to the surface.

The motion of the mobile droplet towards the pinned one is represented in Fig. \ref{fig:2Dmontage} and the distance as a function of time is showed in Fig. \ref{fig:2Dynamics}. The mobile droplet moves at an increasing velocity towards the pinned droplet, until they start to superimpose. The motion then slows down as the droplets align one above the other. The motion is slower than in the unconfined system and can be achieved with an initial separation between the droplets roughly 2.5 times larger than in the three-dimensional case (up to 6 diameters). We also observe that the droplet evaporation is slowed down as it is possible to make the droplet move after more than ten minutes by moving one of the slides. We deduce that the evaporation is reduced by the boundary condition imposed by the plates.  

\subsection{Humidity field}
If the gap between the slides is small compared to their size, the humidity is not a function of the vertical position, and each droplet will create a two-dimensional gradient of humidity, axisymmetric around the droplet, only a function of the horizontal distance to the droplet and of the size of the slides. The vapor diffusion is then axisymmetric and the steady state diffusion equation reduces to:
\begin{equation}
\frac{1}{r}\frac{d}{dr}\left(r \frac{d RH}{dr} \right)=0,
\end{equation}
with $r$ the radial distance to the center of the droplet. With the two conditions at the limit $RH(r=R)=100\%$ and $RH(L)=RH_0$, with $R$ the radius of the droplet and $L$ the typical length of confinement, the solution can be written as:
\begin{equation}
RH=\frac{RH_0-1}{\ln \left( R/L\right)}\ln \left(R/r \right) +1.
\label{eq:RH2D}
\end{equation}
We then expect the humidity field should then be shallower and extend farther away than in the three dimensional case for typical plate size.\\

\begin{figure}
\includegraphics[width=\columnwidth]{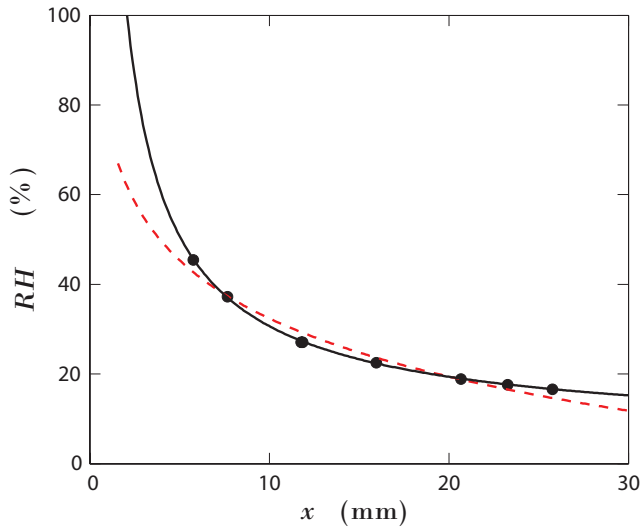}
\caption{Humidity as a function of the distance to a 2 $\mu$L water droplet in two-dimensional confinement. Dots: measurements, doted line: logarithmic fit, plain line: power law fit. \label{fig:2Dgrad}. The origin is the center of the water droplet.}
\end{figure}

We measured the humidity as a function of the distance to the water droplet placed alone in the confinement, in the direction of longer length of the confinement (Fig. \ref{fig:2Dgrad}). We observe that the humidity field is not perfectly fit by Eq. \ref{eq:RH2D}, but that it is indeed smoother and extends further away from the droplet than in the three dimensional case. A better fit of the data is obtained by a power law fit: $RH= \frac{A}{r^\alpha}+B$, with $\alpha=0.9$.

\subsection{Dynamics} 
\begin{figure}
\includegraphics[width=\columnwidth]{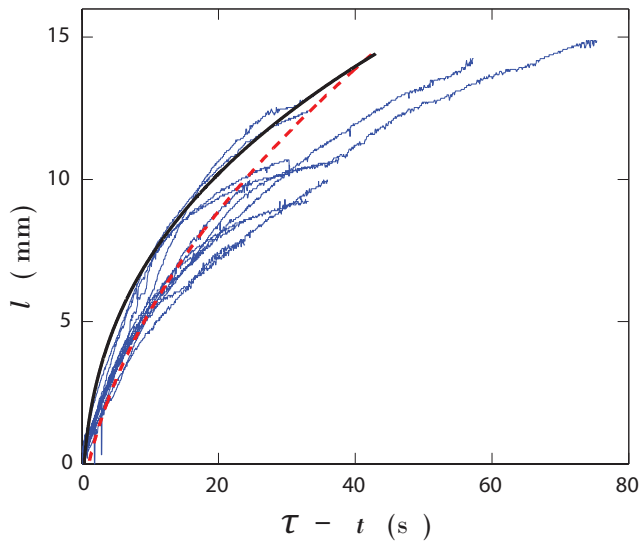}
\caption{Distance between a pure water droplet and a 10$\%$PG droplet as a function of $\tau-t$ in two dimensional confinement. Thin lines: experiments, dashed line: prediction based on a logarithmic fit of the gradient, solid line: prediction based on a power law fit of the gradient. \label{fig:2Dynamics}}
\end{figure}

Fig. \ref{fig:2Dynamics} shows the distance between the droplets as a function of time, before they superimpose upon each other. Using the fits of the humidity field presented in the previous section, we numerically calculated the force between the droplets, integrated to get $l(t)$ and compared to the experimental data. Using values of the humidity field from the logarithmic fit we capture the motion at $l$ smaller than one to two diameters. Using the best fit of the humidity field (power law fit), the model nicely captures the experimental data for $l$ up to 5 diameters. 
In conclusion, in 2D confinement, knowing the humidity field imposed on one droplet, we can predict its motion. Moreover, the confinement of the droplet modifies the humidity field created by one droplet, and allows for a longer reach and auto-alignment of two droplets. The tested confinement did not create a perfectly 2D gradient. One remark about the design: when one droplet is placed in the rectangular confinement, it naturally moves towards the center in the short length direction, because this is where it evaporates less. When we studied attraction between two droplets, the motion happened in the long length direction, so that the mobile droplet was not subject to this additional force. Note that the model assumes that the humidity released by one droplet is not modifying the humidity field from the other, which is physically not the case, and this approximation may not be valid in a stronger confinement.

\section{One-dimensional confinement}
\subsection{Purpose and experimental setup}
Following experiments done in confinement between two plates, we wanted to place a droplet in a constant external gradient of humidity. A constant gradient could be obtained if the vapor was only able to diffuse in one direction with constant humidity values at the boundaries. We built a tunnel 75 mm long ($x$ direction, 2 mm high, and 7 mm wide ($y$ direction) as represented in Fig. \ref{fig: 3D:2D:1D} (d). The tunnel is placed in the controlled humidity box at 10$\%$ RH, and an attractive water puddle 10 mm long by 5 mm wide is placed at one exit of the tunnel, in order to impose constant $RH$ at both ends. Without any external gradient, the droplets would naturally move orthogonally to the tunnel direction, towards one of the vertical walls of the tunnel. To prevent the droplets from contacting the side walls, only the central part of the bottom slide is treated to be completely wetting (\textit{Materials and methods}).

\begin{figure}
\includegraphics[width=\columnwidth]{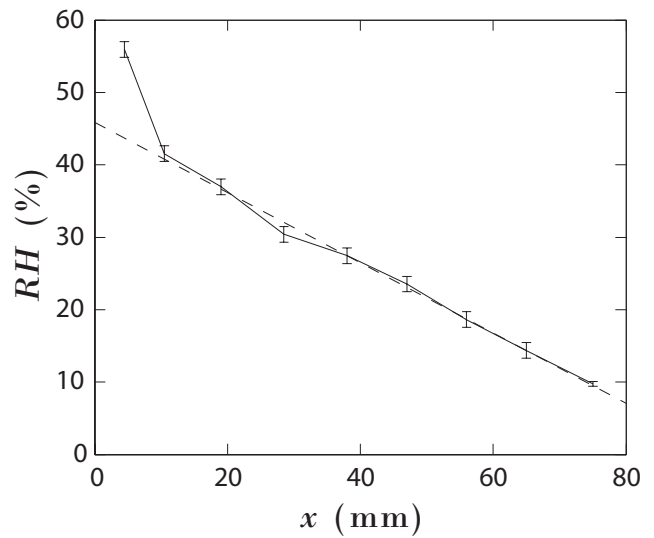}
\caption{Humidity field in the tunnel, with one end exposed to dry air ($x=75$ mm), and one end exposed to a large puddle of water ($x=0$). The error bars show the standard error. \label{fig:1DRHsensor}}
\end{figure}

Fig. \ref{fig:1DRHsensor} shows the humidity field along the tunnel, when we impose the conditions at the boundaries, but no droplet is placed in the tunnel. On the central part of the tunnel, about a centimeter away from both ends, the humidity gradient is constant as expected. Reducing the humidity in the box and a proper placement of the water puddle at one end allows tuning of the constant humidity gradient value.

In a constant gradient of humidity, the force of attraction on the droplet obtained from Eq. \ref{equ:Fattract} reduces to:
\begin{equation}
F_a=\pi \gamma m K R^2
\label{eq:1Dmotion}
\end{equation}
with $K$ the value of the humidity gradient.

\begin{figure}
\includegraphics[width=\columnwidth]{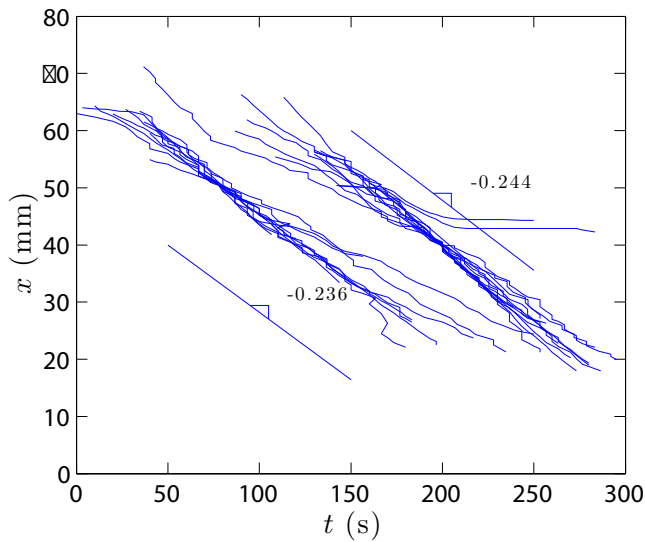}
\caption{Position of the droplet into the tunnel as a function of time. The time is arbitrary so that in all trajectories droplets of the same volume are placed together, left $V=1\mu$L, right $V=0.5 \mu$L. Some trajectories where the droplet pins on a macroscopic defect are intentionally presented. \label{fig:tunnel_positions}}
\end{figure}

We now place a mobile droplet at the dry end of the tunnel and observe its motion in Fig. \ref{fig:tunnel_positions}. We plot the position of droplets of 0.5 $\mu$L and 1 $\mu$L along the tunnel, with $x$ the distance from one end. We observe that the droplets move at a quasi constant velocity from one side of the tunnel to the other. We compare the velocity of the droplet with the value derived from Eq. \ref{eq:1Dmotion}, and find two contradictions. The theoretical velocity of $V=0.5$ $\mu$L droplets is $U=0.026$ mm.s$^{-1}$ and $U=0.033$ mm.s$^{-1}$ for one microliter droplets ($m=0.00025$, $K=0.485$, for $V=0.5$ $\mu$L droplets: $C=0.505$ mm.s$^{-1}\mu N^{-1}$; for $V=1$ $\mu$L droplets: $C=0.391$ mm.s$^{-1}\mu N^{-1}$). First, the velocity of the droplet is about ten times larger than expected. Second, Eq. \ref{eq:1Dmotion} predicts that the driving force scales with $R^2$, and the drag force on a droplet scales with $R$ so that we would expect the velocity of the droplet to scale with $R$, which is not observed. 

\subsection{Self induction}
\begin{figure}
\includegraphics[width=\columnwidth]{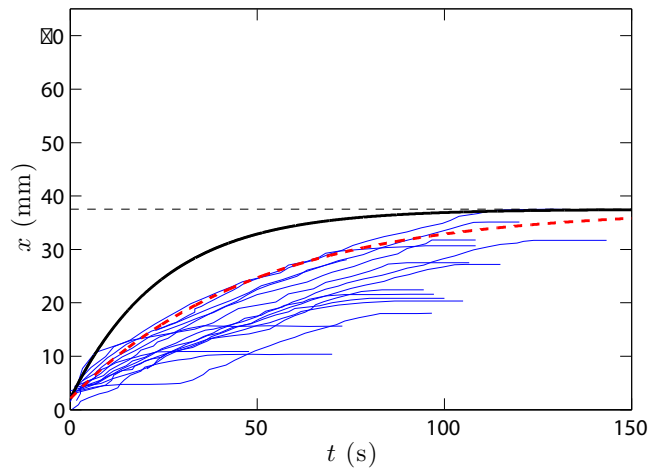}
\caption{Position of droplets of 1 $\mu$L in the tunnel as a function of time, without any external gradient. $x$ represent the distance to the entrance of the tunnel. The solid line represents the numerical prediction, the dashed line the numerical prediction with an attraction force divided by 2. \label{fig:tunnel_self}}
\end{figure}

Trying to understand why the droplets were moving faster than expected in a ¬1D¬ confinement, we made the following control: we placed a droplet in the tunnel without any external gradient, and observed that it moved towards the center of the tunnel (in the $x$ direction). Fig. \ref{fig:tunnel_self} shows the position as a function of time of several droplets of 1 $\mu$L in similar conditions.  We observe that the droplet final position weakly varies between experiments, but the droplets always move towards the center. A physical explanation of this observation is that the droplet itself is creating a gradient of humidity inside the tunnel, that makes it move, by "self-induction". If the gradient is one-dimensional, it is constant on the left (right) of the droplet, of magnitude $d RH/ d x= (1-RH_0)/L_l$ ($d RH/ d x=-(1-RH_0)/L_r$), with $L_l$ ($L_r$) the distance from the droplet to the left (right) end of the tunnel. The magnitude of the gradient determines the rate of evaporation $\Phi$ of the droplet on each side through the equation of diffusion ($\Phi=-1/D \times  d RH/ d x$) with $D$ the diffusivity of water vapor into air. A reduced evaporation means a smaller contact angle of the droplet. If the droplet is on the left of the tunnel, its right side is exposed to a weaker gradient and thus has a smaller contact angle. As a result the droplet moves from left to right, until the gradient from both sides are equal when the droplet centers itself.

\subsection{Model}

To estimate the driving force, as in the previous geometries, we don't have direct access to the concentration of the droplet as a function of the humidity gradient. We measure the contact angle of the droplets when they are placed at the center of tunnels of lengths $L_c=$75, 50 and 25 mm. In this range we observe that $\cos \theta$ increases approximatively linearly with the length of the confining tunnel. We now estimate the force this induces on a moving droplet by assuming that the left side (right side) of the droplet at the distance $L_l$ ($L_r$) from the exit has the contact angle it would have placed in a tunnel of length $L_c=2L_l$ ($L_c=2L_r$). The force one side of the droplet feels is $f=\pm \pi/2 R \gamma \cos\theta$. The global force acting on the droplet is then:
\begin{equation}
F_a=\frac{\pi}{2} \gamma R (\cos \theta_l-\cos \theta_r)
\end{equation}
with $\cos \theta_l$ ($\cos \theta_r$) linear functions of $L_l$ ($L_r$). 

Equating $F_a$ to the drag force we get the theoretical dynamic of the droplets, compared to measurements in Fig. \ref{fig:tunnel_self}. The model captures the essence of the observed dynamic, with a decreasing velocity as the droplets move towards the centre of the tunnel, and eventually predicts a centering of the droplet. Since the continuous variation of the humidity and thus the continuous variation of the contact angle from the left to the right side of the droplet is ignored, the model over-estimates the force. When dividing the estimated force by two, we get a better agreement between the model and experiments. \\

We showed that when the droplet evaporation is modified by confinement, the droplet can self-induce its own motion as it creates a gradient of humidity by itself. This makes it difficult to place a droplet in a linear gradient of humidity as a linear gradient may not be obtained other than by strong confinement, in which the droplet modifies the gradient we try to impose. 

When the droplet is placed in an initially linear gradient in strong confinement, it moves following the gradient, but at velocities larger than expected under the only action of the gradient. To complete this observation, we placed droplets in a less confining tunnel 25 mm wide, 2 mm high and 10 mm long, where we observed that the droplets still moved faster than estimated from the externally imposed gradient force balance, with an estimated force approximatively 25$\%$ smaller than the observed one.

As the glass slides we use are not perfectly clean, the drops pin on macroscopic defects from time to time, similarly to sessile droplets (see Fig. \ref{fig:tunnel_positions} and Fig. \ref{fig:tunnel_self}), with a mechanism that remains to be studied.

\section*{Conclusion}

Two-component droplets of the right liquid mixture do not spread on a high energy surface, but form droplets of constant radius. Contrary to typical sessile droplets, these droplets present a minimal pinning force allowing for motion under minute forces. It allows the droplets to move under the influence of external vapor gradients, imposed by neighboring droplets or from self-generated humidity gradients in confined geometries. Two droplets placed at two to three diameters in an unconfined space will attract each other. The droplets move in the direction of the larger vapor concentration the other is producing by evaporation. The motion is reminiscent to chemotaxy of cells sensing and following chemical cues. The dynamics are well captured by a simple model, assuming that the humidity fields of each droplet can be superimposed. The effect of a vertical wall on a droplet is identical to that of an image droplet with respect to the wall, which shows that first, the image theory can be applied here, and second, confinement can put one droplet alone in motion in a controlled direction.

% Corrected by Nate
%The motion is reminiscent to chemotaxy of cells following chemical cues. The dynamics are well captured by a simple model, assuming that the humidity fields of each droplet can be superimposed. The effect of a vertical wall on a droplet is identical to the one of an image droplet as respect to the wall, what shows that first: the image theory can be applied here; second: confinement can put one droplet alone in motion in a controlled direction. Such sensing of a boundary through the enrichment of the concentration of a  chemical species due to limited diffusion by the boundary is similar to echolocation\cite{griffin1958sensitivity} and even more to active electrolocation \cite{albert2005electroreception} in animals,  but to our knowledge there is no description of the use of a field of diffusive chemical for self-location in the biological realm.

Such sensing of a boundary through the enrichment of the concentration of a chemical species due to limited diffusion by the boundary is similar to echolocation \cite{griffin1958sensitivity, au2007echolocation} and even more to active electrolocation \cite{albert2005electroreception, howard2012perceiving} by animals. 

%Many clues indicate that bacteria could also perceive boundaries by releasing chemical species and sensing their concentration and variation of concentration due to a boundary close by \cite{redfield2002quorum, ponsonnet2008local, tuson2013bacteria}. These mechanisms still lack full comprehension. For example, contrary to droplets, bacteria may be able to sense the proximity of a boundary, but may not be large enough to integrate the signal and estimate the direction of the boundary.

It has been suggested that bacteria may also perceive boundaries by releasing a chemical species and detecting the concentration of that species, inferring presence of the boundary by an increase in concentration which could trigger events such as attachment in biofilm formation \cite{redfield2002quorum, ponsonnet2008local, tuson2013bacteria}. However, these mechanisms still lack the full capability of the droplets presented here. Bacteria are thought to be too small to integrate the signal and estimate the direction of the boundary, even if they know a boundary is near \cite{macnab1972gradient, delbruck1972signal}. We are not aware of any natural example where the direction of a boundary can be determined by diffusive sensing, but hypothesize the existence of this mechanism in biology. 
 
Confinement can also modify the dynamics of attraction between a pair of droplets. Confinement modifies the gradient of humidity each droplet applies on the other, making longer range of attraction possible, up to 20 times the diameter of the droplet in large confinement. As confinement reduces the evaporation rate of droplets, it also increases the lifetime of the droplets. Measuring the external humidity gradients in medium and strong confinements, we are able to predict the dynamic of the droplet with a good accuracy, that decreases as the confinement increases because self-induction gets predominant. In conclusion, by changing the geometry of confinement, one can greatly enhance the range and directionality of motion of single or multiple vapor mediated mobile droplets droplets.

\appendix

\section{Materials and methods}
We use two-component mixtures of propylene glycol (PG) and water. The volume ratio of PG over the total volume is noted xx$\%$ PG. The humidity is measured as a percentage of the humidity at saturation, and is noted xx$\%$ RH. The droplets are deposited on clean microscope glass slides with calibrated pipets, and we record their motion with a DSLR camera at 30 frames per second. The glass slides are cleaned with a corona discharge wand (or a plasma oven) directly out of their packaging. The glass slides are initially partially wetting due to contamination from the ambient air. When there is a need to confine the droplet in a restricted portion of the surface of the slide, only this portion is cleaned, applying an acrylic mask on the glass slide during corona treatment. The experiments are conducted in a humidity controlled box. It is made from a 50x40x50 cm plastic box which was modified, adding a top transparent acrylic observation window, a door and two sealed glove access ports. We reduce the humidity inside the box by blowing dry compressed air from the building network. We monitor the humidity with a silicon based probe and are able to reach humidity as low as 10$\%$ RH, with less than 6$\%$ drift per minute. To measure the humidity in confined geometries, the probe is integrated in a flat plastic piece, with the sensor at the height of the top surface, so that the sensor is placed at a position very close to where a mobile droplet would be.

%\appendix[Estimating the Spectral Norm of a Matrix]

%
%This is an example of an appendix without a title.

%\begin{acknowledgments}
%\end{acknowledgments}

\bibliography{bib_Dconf2}
\bibliographystyle{plain}

\end{document}